\documentclass[pre,aps,twocolumn,showpacs,floatfix,superscriptaddress,10pt]{revtex4}

\usepackage{graphicx}
\usepackage{tensor}
\usepackage{amsmath, amsthm, amssymb}

\usepackage{epsfig}
\usepackage{latexsym}
\usepackage{bm}

\newcommand{\be}{\begin{equation}}
\newcommand{\ee}{\end{equation}}
\newcommand{\bea}{\begin{eqnarray}}
\newcommand{\eea}{\end{eqnarray}}

\def\beginwide{
        \end{multicols} \vspace*{-0.5cm} \noindent
        \rule{3.5in}{.1mm}\rule{.1mm}{5mm} \widetext \medskip }
\def\beginwidetop{
        \end{multicols} \vspace*{-0.5cm} \noindent
        \widetext \medskip }
\def\endwide{
        \hspace*{3.35in}~\rule[-5mm]{.1mm}{5mm}\rule{3.5in}{.1mm}
        \begin{multicols}{2} \vspace*{-1.0cm} \noindent }
\def\endwidebottom{
        \begin{multicols}{2} \vspace*{-1.0cm} \noindent }

\begin{document}

\title{Cooperation amongst competing agents  in minority games}
\author{Deepak Dhar}
\email{ddhar@theory.tifr.res.in}
\affiliation{Department of Theoretical 
Physics, Tata Institute of Fundamental Research, Homi Bhabha Road, Mumbai-400005, India.}
\author{V. Sasidevan}
\email{sasi@theory.tifr.res.in}
\affiliation{Department of Theoretical 
Physics, Tata Institute of Fundamental Research, Homi Bhabha Road, Mumbai-400005, India.}
\author{Bikas K. Chakrabarti}
\email{bikask.chakrabarti@saha.ac.in}
\affiliation{Saha Institute of Nuclear Physics, Sector-I, Block AF, Bidhannagar , Kolkata-700064, India.}
\affiliation{Economic Research Unit, Indian Statistical Institute, Kolkata-700108, India.}

\begin{abstract} 
We study a variation of the minority game. There are $N$ agents. Each has to choose between one of two  alternatives everyday, and there is reward to each member of  the smaller group. The agents cannot communicate with each other, but try to guess the choice others will make, based only the past history of number of people choosing the two alternatives.  We describe a simple probabilistic strategy using which the agents acting independently, can still maximize  the average number of people benefitting every day. The strategy leads to a very efficient utilization of resources, and the average deviation from the maximum possible can be made ${\mathcal O}(N^{\epsilon})$, for any $\epsilon >0$.  We also show that a single  agent does not expect to gain by not following the strategy. 

\end{abstract}

\maketitle
\section{ Introduction}
\label{sec1}

The Minority Game(MG) is a particular version of the  El Farol Bar problem. The latter was introduced by  Brian Arthur  as a prototypical model for the complex emergent behavior in a system of many interacting agents  having only incomplete information, and bounded rationality \cite{elfarol}.
This problem is about $N$ agents, who have to repeatedly make  choices between two alternatives, and the winners are those who selected the alternative chosen by fewer agents. 
MG has been studied a lot as  a mathematical model of   learning, adaptation, and co-evolution of agents \cite{mgpapers1, mgpapers2}. An overview and bibliography may be  found in \cite{mgbook, mg1, mg2}. 
The interesting feature of the minority game is that the agents seem to be able to coordinate their actions, without any direct communication with each other, and  the system can self-organize  to a state in which 
the fluctuations in the steady state  are much less than what would be expected if each agent made a random choice.  This is called the efficiency of the markets.

In  a system of $N$ interacting agents, with $N$ odd, the degree of efficiency of the system   may be measured by how close is the average  number of happy agents in the steady state to the maximum possible value $(N-1)/2$. Simulations of MG have shown that typically the difference is of order $N^{1/2}$. The coefficient
of the $N^{1/2}$ depends on details of the model, like how far back in the past the agents look to decide their action, but it can be much less than the value for agents making random choices. The minimum value of the coefficient attained in  several variants of the MG  is about $1/10$ \cite{mg1}.

A variation of the minority game, focussing on the efficient utilization of resources was studied by Chakrabarti et al as the Kolkata Paise Restaurant problem \cite{kpr1, kpr2, kpr3}. In this variation, there are $N$ restaurants, and $N$ agents, and there is rank order amongst  the restaurants. Each restaurant can take only one agent per day, and agents prefer to go to a higher ranked restaurant. In spite of this complication, it was found that  an egalitarian probabilistic strategy exists in which the agents visit restaurants in a cyclic order. Also, the agents can reach this cyclic state in  a short time. 

In this paper, we describe a probabilistic strategy, inspired by the strategy  suggested in \cite{kpr3}, for the  minority games, that is very  simple, but is   more efficient than those previously studied in literature.  In this strategy, the average deviation of the number of people in the minority  from the maximum $(N-1)/2$ can be reduced to be of order $N^{\epsilon}$, for any  $\epsilon > 0$, and the time required to reach this level increases with $N$ only as $\log \log N$. In addition, we show that a game where all agents follow this strategy, is stable against individual cheaters.  

 Our strategy  is an application  of the general win-stay-lose-shift strategy
\cite{nowak}, an adaptation of which to MG was discussed earlier by Reents et al \cite{reents}.  In the latter, the deviation from best possible can be made of order $1$, but the time required grows as $N^{1/2}$.  We are able to get a much faster approach to optimum by using a shift probability that depends on the current distance from optimum. Other probabilistic strategies for minority games have also been discussed in literature \cite{thermal1, thermal2, xie}, and in some cases, it has been noted that, they can perform better than the deterministic ones \cite {sornette}. While our strategy seems more or less obvious, we did not find it discussed in the literature so far, and it seems worthwhile to study it quantitatively.

The plan of the paper is as follows: in section \ref{sec2}, we define the rules of the game precisely  and  argue that the strategy defined leads to a very efficient use of resources. In section \ref{sec3}, we show that individual agents have no incentive to cheat, if every body else follows the same strategy. Section \ref{sec4} contains
the results of our simulations of the model, and \ref{sec5} contains some concluding remarks.

\section{Definition of the model}
\label{sec2}

The model we consider is  a variation of the   El Farol Bar problem. We consider a small city with exactly two restaurants. There are $N$ people  in the city, called agents, each of whom goes for dinner  every evening to one of the two restaurants. The prices and quality of food is quite similar in both, and the only thing that governs the choice of agents about which restaurant they go to on  a particular day is that the quality of service is worse if the restaurant is crowded. We assume that $N$ is odd, and write $N = 2 M +1$. 
The restaurant is said to be crowded on a particular day, if the number of people turning up to eat there that day exceeds $M$. An agent is happy if he goes to  a restaurant that is uncrowded, and will be said to have  a payoff $1$. He turns up at a crowded restaurant, his payoff is $0$.  Once the choice of which restaurant to go to is made, an agent cannot change it for that day.

The agents can not communicate with each other in any way directly in deciding which restaurant to go to.
However, each of them has available to him/her the entire earlier history of how many people chose to go to the first restaurant (call it A), on any earlier day.  Let us denote the number of agents turning up at A on the $t-$th day by $M - \Delta(t)$. Then the number of agents turning up at the Restaurant B are $M + \Delta(t) +1$.  At the end of day $t$, the value of $\Delta(t)$ is made public, and is known to all the agents.   Using the information $\{\Delta(t')\}$, for $t' =  1,2 ...t$, the agents try to guess
the choice that other customers who share the same public knowledge will make, and  decide which restaurant to go to on the day $(t+1)$, and  try to optimize their
payoff. 

In the standard MG, the public information is not the value of $\Delta(t)$, but  only whether it was negative or not \cite{mgpapers1, mgpapers2}. [ In contrast, in our model, the agents have better quality of information, and this difference is important.]  Also, in MG each agent has a finite  set of strategies available to him/her, which  uses only the history $\{\Delta(t)\}$ for  $m$ previous days, where $m$ is a fixed non-negative integer. Each strategy  is deterministic: for a given history, it  tells which restaurant agent should go to.  While the 
agent has more than one strategy available to him/her,  he chooses the strategy that has the best `performance score' in the recent past. There is no probabilistic component in the choice of any agent. For  a given history, the future choices of all agents for all subsequent days are fully determined.

In the problem we study here, we allow agents to have probabilistic strategies.  For a given history $\{\Delta(t)\}$, a strategy will specify a probability $p$ with which he should  go to restaurant A. Another important difference from the MG's is that we allow the strategy to depend explicitly  on the payoffs received in the $m$ previous days.  In MG, the strategy does not explicitly involve previous payoffs. The payoff only affect the outcome indirectly, through the performance scores that determine which strategy is used by the agent.

The  simplest case  corresponds to $m=0$. In this case, an agent has no information. His probabilistic strategy is to make  a random choice of which restaurant to go to, with equal probability.  In this case, the probability that $r$ people show up at Restaurant A is clearly,  is  
\begin{equation}
{\rm Prob}(r) = 
{\begin{pmatrix}  N\\ r \end{pmatrix}} 2^{-N}
\end{equation}

The expectation value of $r$ is $N/2$, and for large $N$, the distribution is nearly gaussian, with a width proportional to $\sqrt{N}$. 
 We can measure the inefficiency  of the system by a parameter $\eta$ defined as 
\begin{equation}
\eta = \lim_{N \rightarrow \infty} \frac{4}{N} \langle (r -N/2)^2 \rangle
\end{equation}
where $\langle ~\rangle$ denotes averaging over a long time evolution, and over different initial conditions.

The normalization has been chosen, so that the inefficiency parameter $\eta$ of the system with agents using his /her choice randomly is $1$.

We now describe a simple $m=1$ probabilistic strategy, that  gives a highly efficient system, where inefficiency
parameter can be made of order $(1/N^{1 - \epsilon})$, for any $\epsilon >0$. 

The strategy is defined as  follows: At $t=0$, each agent chooses one of the two restaurants with probability $1/2$ each, independently of others. At any subsequent time $t+1$,  each agent follows the same simple strategy : If at time $t$, he found himself in the minority, he chooses the same restaurant as at time $t$. If he found himself in the majority, and the number of people visiting the same restaurant as him was $M + \Delta(t) +1$, with $\Delta(t) \geq 0$,  he changes his choice with a small probability $p$, and sticks to earlier choice with probability $1 -p$, independent of other agents. The  value of $p$ depends only on  $\Delta(t)$. It is approximately equal to $ \Delta /M$ for $\Delta >0$. The precise dependence of $p$ on $\Delta$ is discussed  later in the paper.  

For large $M$, the number of people changing their choice is distributed according to the Poisson distribution,
with mean approximately equal to $\Delta$, and width varying as $\sqrt{\Delta(t)}$. Thus we have the approximate recursion $ \Delta (t+1) \approx \sqrt{\Delta(t)}$, for $\Delta(t) \gg 1$. This shows  that within  a time of order $\log \log  N$, the magnitude of $\Delta$ will become of ${\cal O}(1)$, and then remain of order $1$.

\section{Stability against individual cheaters}
\label{sec3}

In the previous section, we have shown that if all the agents follow the proposed common strategy,  the social inefficiency of the system is considerably reduced.  However, selfish agents may not do what is expected of them for social good, and   act differently, if it gives them  profit.  In this section, we show that if all the other people are following the common strategy outlined above, there is a specially selected 
value of $p$, for each $\Delta >0$, such that if the other agents follow the strategy with this value of $p$,  a single
individual gains no advantage by cheating. 

The emergence of effective cooperation amongst selfish agents in our problem may seem rather paradoxical at first. After all, the main point of MG is that agents gain by differentiating, and not following the same strategy. If rational agents know that they cannot improve their immediate individual gain by cheating, they would then try  to maximize their individual long-term payoff. This they can do, if they follow the same  common strategy. {\it This cooperative strategy is beneficial for everybody in the long run, and deviating from it has no advantage}. This is the reason for the emergent cooperation between agents in our model.

Let us consider any particular day  $t$. Let the number of people who showed up in the restaurant A be $M -\Delta(t)$.  We may assume $\Delta(t) \geq 0$ , without loss of generality. 

We consider first the case $\Delta >0$. We consider a particular agent Alice, who went to A on the $t$-th day, and found herself in the happy situation of being in the minority. Alice assumes that all other agents  follow the strategy.
Then, all other agents who went to A will go to it again on day $(t+1)$. There are $M + \Delta +1$ agents that went to B. Each of these agents will change his/her  choice with probability $p$. Let $r$ be the number of agents that actually change their choice at time $(t+1)$.   Then, $r$ is  a random variable, with a distribution given by
\begin{equation}
{\rm Prob}_{p} (r) = \dbinom{M + \Delta +1}{r}  p^{r} (1 - p)^{M+ \Delta +1 -r}	
\end{equation}

For $M \gg 1$, this distribution tends to the Poisson distribution with parameter $\lambda = p ( M + \Delta +1)$, given by
 \begin{equation}
{\rm Prob}_{\lambda}(r) = \lambda^r e^{ -\lambda }/ r!	
\end{equation}

If Alice chooses to go  to A the next day, she will be in the winning position, if $r \leq \Delta$.
Hence her expected payoff $EP(Alice|stay)$, if she chooses to stay with her present choice is
\begin{equation}
EP(Alice|stay) = \sum_{r=0}^{\Delta} {\rm Prob}_{p} (r)
\end{equation}

If, on the other hand, Alice would switch her choice, she would win if $ r \geq \Delta+2$. Hence, we have
her  expected payoff $EP(Alice|switch) $ if she chooses to switch, given by
\begin{equation}
EP(Alice|switch) = \sum_{r=\Delta+2}^{\infty} {\rm Prob}_{p} (r)
\end{equation}

For Alice to have no incentive to cheat, we must have
\begin{equation}
EP(Alice|stay) \geq EP(Alice|switch).
\label{eq:Alice}
\end{equation}

Now consider the agent Bob, who went to B on day $t$.  He also assumes that all other people will follow the
strategy: those who went to A will stick to their choice, and  those who went to B switch their choice with probability $p$.  There are $ M + \Delta$ other persons who went to B. If Bob chooses to cheat, and decide to stay put, without using a random number generator, the number of agents  switching would be a random number $	\tilde{r}$, with a distribution given by
\begin{equation}
{\rm Prob}'_{p}(\tilde{r}) = \dbinom{M + \Delta}{\tilde{r}}  p^{ \tilde{r}} (1 - p)^{M+ \Delta  - \tilde{r}}	
\end{equation}
He would be in the minority, if $\tilde{r} \ge \Delta +1$. Thus, if he chooses to stay, we have his expected payoff $EP(Bob|stay)$ given by
\begin{equation}
EP(Bob|stay) = \sum_{\tilde{r}=\Delta+1}^{\infty} {\rm Prob}'_{p} (\tilde{r})
\end{equation}

On the other hand, if Bob decide to switch his choice, he would win if $ \tilde{r} \leq \Delta-1$. In that case, his expected payoff $EP(Bob|switch)$ is given by
\begin{equation}
EP(Bob|switch) = \sum_{\tilde{r}=0}^{\Delta -1} {\rm Prob}'_{p} (\tilde{r})
\end{equation}

We choose the value of $p$ to make these equal. Thus the equation determining $p$, for a given $\Delta$ and $N$ is 
\begin{equation}
EP(Bob|stay) = EP(Bob|switch) 
\label{eq:Bob}
\end{equation}

If the above condition is satisfied, Bob can choose to stay, or switch, and his expected payoff is the same. More generally, he can choose to switch with a probability $\alpha$, and his payoff is independent of $\alpha$. In that case, what is the optimum value of $\alpha$ for Bob?  One has to bring in a different optimization rule 
to decide this, and it seems reasonable that Bob would choose a value that  optimizes his long-time average payoff, (which is the same for any other agent), and hence choose the value $p$.

In the limit of $M \gg \Delta$,  eq. (\ref{eq:Bob}) simplifies, as the dependence on $M$ drops out, and we get a simple equation determining  the  dependence of the Poisson  parameter $\lambda$ on $\Delta$.
 Then, Eq. (\ref{eq:Bob}) becomes
\begin{equation}
\sum_{r=0}^{\Delta -1}  \frac{ \lambda^r }{ r!}	e^{-\lambda}= \sum_{r= \Delta +1}^{\infty} \frac{ \lambda^r} { r!	}e^{-\lambda}
\label{eq:12}
\end{equation}

This equation may be rewritten, avoiding the infinite summation,  as 
\begin{equation}
2 \sum_{r=0}^{\Delta -1} \frac{  \lambda^r e^{-\lambda }}{ r!}	= 1 - \frac{\lambda^\Delta e^{- \lambda } }{ \Delta!}	
\label{eq:lambda}
\end{equation}

It is easy to see that Eq.(\ref{eq:lambda}) implies that Eq.(\ref{eq:Alice}) is also satisfied. 
For the sake of simplicity, we will only consider this limit of large $M$ in the following. The extention to finite $M$  presents no special difficulties.

Thus, for any given value of $\Delta > 0$, the optimum value of $\lambda$ is determined by solution of Eq. (\ref{eq:lambda}). This equation is easily solved. The resulting values of $\lambda$ for different $\Delta$ are shown in Table \ref{table1}. For large $\Delta$, we show in the  Appendix that ($\lambda - \Delta $) tends to $ 1/6$. 

\begin{figure}

\begin{center} \includegraphics[width=8.0cm]{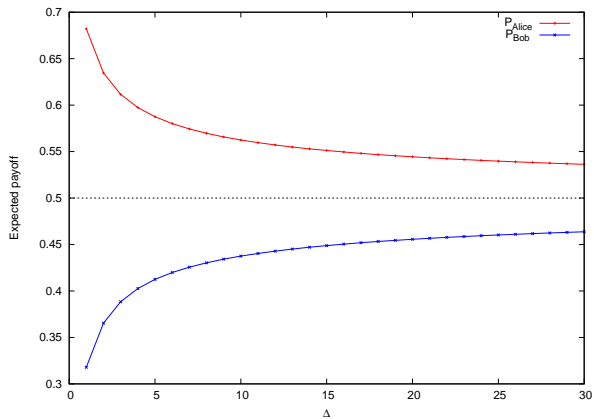} \caption{Variation of expected payoff for the next day of an agent in Restaurant A ($P_{Alice}$) and  Restaurant B ($P_{Bob}$) with $\Delta$.}
\label{fig1}
\end{center}
\end{figure}
We note that the values  of $\lambda$ do not have to  be broadcast to the agents by any central authority. Each individual rational agents will be able to deduce them as optimal, without any need to communicate with others.
Fig. \ref{fig1} shows the variation of the expected payoff for the next day of Alice and Bob with $\Delta$. As expected we can see that for large values of $\Delta$, the expected payoff of an agent in either restaurant tend to the value $1/2$. Alice's payoff is a bit bigger than $1/2$, but this advantage is short-lived. Also, Bob cannot utilize this predictability of the system, as an attempt to switch by him change the outcome with finite probability. 

 \begin{table}[ht]
 \caption{}
\centering
\begin{tabular}{| c |  r | c | r |}
\hline\hline
$\Delta$ & \multicolumn{1}{c |}{$\lambda$} & $\Delta$ & \multicolumn{1}{c |}{$\lambda$
}\\[0.5ex]
\hline
1 & 1.14619 &8 & 8.16393 \\
2 & 2.15592 & 9 & 9.16423 \\
3 & 3.15942 & 10 & 10.16448\\
4 & 4.16121 &  20 & 20.16557\\
5 & 5.16229 &30 & 30.16594\\
6 & 6.16302 & 40 & 40.16612\\
7 & 7.16354 & 50 & 50.16623\\[1ex]
\hline
\end{tabular}
\label{table1}  

\end{table}

Now, we consider the case $\Delta =0$. In this case, restaurant A has exactly $M$, and B has $M +1$ people.   We now show that  there is no optimum value of $\lambda$ in this case. 

A naive extention of the strategy for $\Delta >0$ to this case would be that Alice does not switch. But then, if there is a nonzero $\lambda$, and agents from B switch to A,  Bob has an  incentive to cheat, as if he goes to A, he would be sure to be  in the majority. If he cheats, and stays back, but at least one other people leave
from B to A ( which occurs with nonzero probability for any non-zero $\lambda$), he has some chance to be on the winning side. 

Clearly, $\lambda =0$ is not a viable strategy, as then nobody switches, and the state at day $(t +1)$ is same as on day $t$. And same situation is met again.  While this is a solution which minimizes wastage of resources, and is `socially efficient',  this is clearly a very unfair state of affairs, where a subset of people are privileged, and have  payoff $1$ every day, and another set has no chance of 
any payoff. 

Consider the possible strategy  that  in this case, all people who went to A switch with probability $\lambda'/M$, and all who went to B switch with probability $\lambda''/(M+1)$, with both $\lambda'$ and $\lambda''$ non-zero.
Let  $r'$ and $r''$ be the random variables denoting  the number of people switching sides from A to B, and from B to A respectively.  Then, $r'$ and $r''$ are  Poisson-distributed independent random variables with mean $\lambda'$ and $\lambda"$ respectively.  Repeating the analysis above, we see that the condition that Alice has no incentive to cheat gives the condition
\begin{equation}
{\rm Prob}( r' <  r''-2 ) ={\rm Prob}( r' \geq  r'')
\label{eq:d0:1}
\end{equation}
Similarly, for the absence of incentive to cheat for Bob, we should have 
\begin{equation}
{\rm Prob}( r' <  r''-1 ) ={\rm Prob}( r' \geq  r''+1)
\label{eq:d0:2}
\end{equation}
It is easy to see that Eq. (\ref{eq:d0:1}) and Eq. (\ref{eq:d0:2}) are mutually inconsistent, as the LHS of
the former is strictly less than the LHS of the latter, and for RHS it is the opposite. Thus, we cannot find 
nonzero finite values $\lambda'$ and $\lambda''$, which will give  a stable strategy against individuals cheating.

Thus, if we reach $\Delta =0$, it is not clear what any agent should do.  We note that in this case, though Bob does not expect to gain anything on the next day by switching, he would still like to do that to upset the status quo, and 
improve his chance of winning the day after.  Of course, as Alice realizes that some people from B are likely to switch, she would like to switch as well.
A simple strategy is that in this case, all agents irrespective of whether they were in minority or not on day $t$,   switch   the next day  with a proability $M^{\epsilon -1}$, where $\epsilon$ is a real number $0 \leq \epsilon \leq 1$. This corresponds to 
both $\lambda'$ and $\lambda''$ very large, of order $M^{\epsilon}$. We shall refer to this step as a major resetting event.

For a given value of $\epsilon$, the value of $|\Delta|$ just after resetting is of order $M^{\epsilon/2}$. Then 
it lakes time of order $\log \log{M}$ to reach the value $\Delta =0$. Then the maximum contribution to the mean efficiency parameter comes from the major resetting events, and it is easy to see that the mean inefficiency parameter 
would vary as $M^{\epsilon -1}/\log \log {M}$.  Then, for more efficiency, we should keep $\epsilon$  small.

\section{Monte Carlo simulations}
\label{sec4}
We have studied the time evolution of  a set of $N$ agents using  this strategy using Monte Carlo simulations, with $N=2001$. If the restaurant with greater attendance has $M + 1 + \Delta$  agents on a given day, with $\Delta > 0$, the next day
each of them switches his/her choice with probability $\lambda(\Delta)/( M + \Delta +1)$, and the agents in the minority restaurant stick to their choice. If there are exactly $M +1$ agents in the majority restaurant, all agents switch their restaurant with a probability $ 1/(2  M^{1 -\epsilon})$.

\begin{figure}

\begin{center} \includegraphics[width=8.0cm]{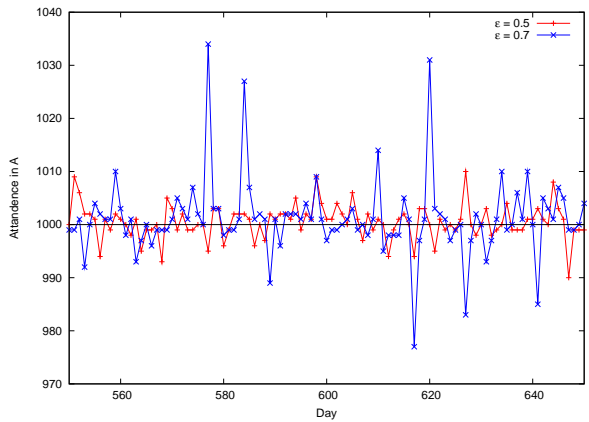} \caption{ A typical evolution of a system  of $2001$ agents for two different choices of the parameter $\epsilon$ = $ 0.5$ and $0.7$.  The large deviations correspond to  major events (see text). } 
\label{fig2}
\end{center}

\end{figure}

The result of a typical evolution is shown in Fig. \ref{fig2}, for two choices of $\epsilon$: $0.5$ and $0.7$. We see that the majority restaurant changes quite frequently.  The large peaks in $|\Delta|$ correspond to resettings of the system, and clearly, their magnitude decreases if $\epsilon$ is decreased. There is very little memory of the location of majority
restaurant in the system. To be specific, let $S(t)$ is $+1$ if the minority restaurant is A in the $t$-th step, and $-1$ if it is B. Then the autocorrelation function  $\langle S(t) S(t +\tau)\rangle$ decays exponentially with $\tau$, approximately as $\exp( - K \tau)$. The value of $K$ depends on $\epsilon$, but is about $2$, and the correlation is negligible for $\tau> 3$.

\begin{figure}
\begin{center} \includegraphics[width=8.0cm]{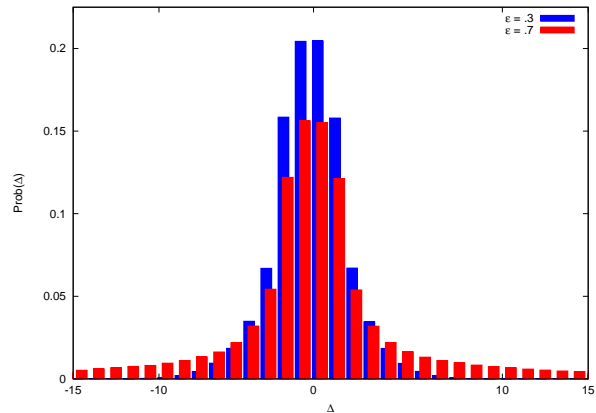} 
\caption{Probability distribution of $\Delta$ in the steady state for $\epsilon = .3,.7$ obtained by evolving $N = 2001$ agents for $10^{6}$ time steps. The red bars have been shifted a bit to the right for visual clarity.} 
\label{fig3}
\end{center}
\end{figure}

\begin{figure}

\begin{center} \includegraphics[width=8.0cm]{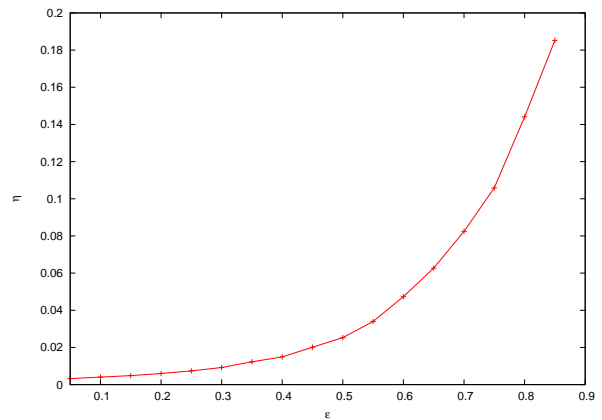} \caption{Variation of inefficiency parameter $\eta$ with $\epsilon$, obtained by averaging the evolution of  $N = 2001$ agents for 10000 time steps.} 
\label{fig4}
\end{center}
\end{figure}

Fig. \ref{fig3} shows the probability distribution of $\Delta$ in the steady state for two different values of $\epsilon$. Fig. \ref{fig4} gives a plot of the inefficiency parameter $\eta$ as a function of $\epsilon$. In each case, the  estimate of $\eta$  was obtained using a single evolution of the system for $10000$ time steps. The fractional error of estimate is less than the size of symbols used.

\begin{figure}
\begin{center} \includegraphics[width=8.0cm]{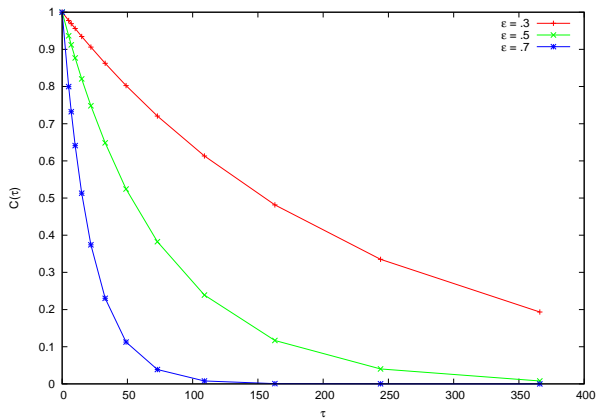} \caption{$C(\tau)$ as a function of $\tau$ for $\epsilon = .3, .5$ and $.7$. Each data point is obtained by averaging over 10000 simulation steps.Total number of agents is N = 2001.} 
\label{fig5}
\end{center}

\end{figure}

We define $A_i(t)$ equal to $+1$ if the $i$-th agent was in the restaurant A at time $t$, and $-1$ otherwise.   We define the auto-correlation function of the $A$-variables in the steady state  as 
\begin{equation}
C(\tau) = \frac{1}{N} \sum_i  \langle A_i(t) A_i(t +\tau) \rangle
\end{equation}
In Fig. \ref{fig5}, we have shown the variation of $C(\tau)$ with $\tau$. We see that this function has  a large amount of persistence. This is related  to the fact that only  a small fraction of agents actually switch their choice at any time step. Clearly, the persistence time is larger for smaller $\epsilon$.

\section{Discussion}
\label{sec5}

In our analysis of the strategy discussed, we assumed that whenever the system reaches a state in which one restaurant has exactly $M$ agents,  it is not possible to find a strategy for reaching a nearby state, with only a few  agents switching, and the system undergoes a major resetting.  
However, consider  a situation where because of shared common history, the agents  agree to a convention that
if such a state is reached, it continues for $T$ more days without change, as it is socially efficient, and on the $(T+1)$-th  day, the major resetting occurs. The rationale  for such a choice would be that  all agents recognize that this state has overall maximum social benefit, and in the long run, any agent would spend equal amount of time  in the privileged  class. Clearly, for realistic modelling, $T$ should not be too large. It has to be significantly less than the expected lifetime of an agent. 

The number of consecutive days when $\Delta$ is nonzero is of order $\log \log N$, and then for $T$ consecutive days $\Delta $  remains zero. Then,  the volatility $\eta$ in such a strategy is given by

\begin{equation}
\eta \simeq  \frac{ K_1 N^{\epsilon -1}}{ T + K_2 \log \log N}
\end{equation}
where $K_1$ and $K_2$ are some constants.

This conclusion is not very surprising. A society that has a larger value of $T$ has more overall social benefit than one with a shorter value. However, agents have to look for something other than  payoff on the next day to realize this, and one  needs to go beyond  myopic strategies that maximize the payoff on the next day. 
An interesting question is what strategies would emerge if the agents try to  maximize the sum of their expected  payoffs in the next $n$ days for $n > 1$.

Generalization of this strategy to the Kolkata Paise Restaurant problem is straight forward.  The strategy is as follows:
If an agent was fed at restaurant of rank $k$ at time step $t$, he goes to restaurant of rank  $k-1$ at time $t+1$. If he found no food at  time step $t$,  He picks at random one restaurant, out  of the restaurants that had no customer at step $t$. If the picked restaurant has rank $k'$, he goes to the restaurant with rank $k'-1$. 
Then, the average time required to reach a cyclic state is of order $\log N$. And in the cyclic state, each agent gets to sample all the restaurants. The strategy can be made robust against cheaters, if we make the additional rule that if more than one customer shows up at the restaurant of rank $k$, preference is given to the customer who was served at rank $(k+1)$ restaurant the previous day.

An interesting question is the effect of heterogeneity in agents, as far as the value of $\epsilon$ is concerned.
There may be impatient agents that do not want to wait, and switch with probability $1/2$ as  soon as the value $\Delta =0$ is reached. If the number of such agents is $N^{a}$, with $ a <1$, it is easy to see that the final efficiency parameter can not be less than $ N^{a-1}$.  In order to get substantial decease in inefficiency, the number of such agents should be small.

The optimum value of $T$, or of the parameter $\epsilon$ is not decidable within the framework of our model, as one needs to bring in other criteria like fairness or social equality, and  decide  the relative weights of these and social efficiency and the wish to have the next win  quickly to determine the optimum choice. Also their have to be some general shared values amongst the agents to make this possible. Clearly, a discussion of these issues is beyond the scope of our work.

Acknowledgements: We thank Dr. Bill Yeung for a very useful correspondence. The work of DD is supported in part 
by Department of Science and Technology, Government of India by the grant SR/S2/JCB-24/2006.

\appendix*
\section{}
\label{app}

In this Appendix, we discuss the solution of the equation (\ref{eq:12})
\begin{equation}
\sum_{r=0}^{\Delta -1}   f_{\lambda}(r)= \sum_{r= \Delta +1}^{\infty}  f_{\lambda} (r)
\end{equation}
where  $f_{\lambda}(r) = \lambda^r \exp( -\lambda)/ \Gamma(r+1)$, for $r$ not necessarily integer. 
We want to solve for $\lambda$, when $\Delta$ is given to be a large positive integer.  We want to show 
in the limit of large $\Delta$, $\lambda - \Delta$ tends to $1/6$. 

For large $\lambda$, the Poisson distribution tends to  a gaussian centered at $\lambda$, of variance $\lambda$.
If the distribution for large $\lambda$ were fully symmetric about the mean, the solution to the above equation would be 
$\lambda = \Delta$. The fact that difference between these remains finite is due to the residual asymmetry in the
Poisson distribution, for large $\lambda$.

For large $\lambda$, $f_{\lambda} (r)$ is a slowly varying function of its argument.  We add
$f(\Delta)/2$ to both sides of eq. (\ref{eq:12}), and approximate the summation by an integration. Then, the eq. (\ref{eq:12}) can  be approximated by 
\begin{equation}
\int_{0}^{\Delta} f_{\lambda}(r) dr = \int_{\Delta}^{+\infty} f_{\lambda}(r) dr =1/2
\label{eq:12p}
\end{equation}

We have used the  trapezoid rule 
\begin{equation}
 \left[ f(r) + f(r+1) \right]/2 \approx   \int_r^{r+1}   dr' f(r'),
\end{equation}
It can be shown that the discrepancy between Eqs. (\ref{eq:12}) and (\ref{eq:12p})  is at most  of order $(1/\lambda)$.

Then, for large $\lambda$, deviations of $f_{\lambda}(r)$  from the limiting gaussian form   can be 
expanded in inverse half-integer powers of $\lambda$ 
\begin{equation}
f_{\lambda}(r) =  \frac{1}{\sqrt{\lambda}} \phi_0 (x) +  \frac{1}{\lambda} \phi_1 (x) + \ldots.
\label{eq:a1}
\end{equation}
where $x$ is a scaling variable defined by $x = ( r - \lambda)/\sqrt{\lambda}$.
Here $\phi_0(x)$ is the asymptotic gaussian  part of the distribution, as  expected from  the central limit theorem,
and  $\phi_1 (x)$  describes the first correction term.

The  characteristic function for the Poisson distribution ${\tilde \Phi}_{\lambda}(k)$ defined by 
\begin{equation}
{\tilde \Phi}_{\lambda} (k) = \langle e^{i k r} \rangle = \sum_{r=0}^{\infty} e^{i k r }  {\rm Prob}_{\lambda} (r) = \exp \left[  \lambda e^{i k} -   \lambda \right] \\
\nonumber
\end{equation}
\begin{equation}
= \exp \left[ i k \lambda - k^2 \lambda /2 - i k^3 \lambda/6 +.. \right]
\end{equation}

Keeping the terms up to quadratic in $k$ gives  the asymptotic gaussian form of the central limit theorem
\begin{equation}
\phi_0(x) = \frac{1}{\sqrt{2 \pi}} \exp( - x^2/2).
\label{eq:phisx}
\end{equation}
The first order correction to this asymptotic form of ${\tilde \Phi}_{\lambda} (k) $  is given by 
\begin{equation}
\tilde{\phi}_1 (k) = \frac{- i k^3}{6} \exp(-k^2 /2)
\label{eq:phiak}
\end{equation}
which gives on taking inverse Fourier transforms
\begin{equation}
\phi_1 (x) = \frac{1}{6 } \frac{d^3}{dx^3} \phi_0 (x)
\label{eq:aa}
\end{equation}

Substituting the functional forms for $\phi_0 (x)$ and $\phi_1 (x)$ in Eq. (\ref{eq:12p}), we get
\begin{equation}
\int_{-\infty}^{\frac{\Delta - \lambda}{\sqrt{\lambda}} }dx \left[ \phi_0 (x) + \frac{1}{\sqrt{\lambda}} \phi_1(x) \right] = 1/2.
\end{equation}

Now, $\phi_1(x)$ is an odd function of $x$, and is zero for $x =0$. As $\Delta -\lambda$ is small,  in the coefficient of $1/\sqrt{\lambda}$, we can replace the upper limit of the integral by zero.  Thus we write
\begin{equation}
\int_{-\infty}^{(\Delta -\lambda)/\sqrt{\lambda}} \phi_1 (x') dx' \approx \int_{-\infty}^{0} \phi_1 (x') dx'
\label{eq:ab}
\end{equation}

But using eq. (\ref{eq:aa}), we get 
\begin{equation}
\int_{-\infty}^{0} \phi_1 (x') dx' = \frac{1}{6} \frac{d^2}{dx^2} \phi_0 (x){\displaystyle  \vert}_{x=0} = -\phi_0(0)/6
\end{equation}

Substituting in eq. (\ref{eq:ab}), we get 
\begin{equation}
\int_{-\infty}^{(\Delta -\lambda)/\sqrt{\lambda}} \phi_0(x') dx' = 1/2 - \frac{\phi_0 (0)}{6 \sqrt{\lambda}}+{\mathcal O}(1/\lambda)
\end{equation}
 and comparing terms of  order $\lambda^{-1/2}$ we get
\begin{equation}
\lambda - \Delta = 1/6 +{\mathcal O}(\frac{1}{\sqrt{\lambda}}).
\end{equation}

\end{document}